\documentclass[twoside,fleqn]{ActaStyle}
\usepackage[dvips]{graphics}
\usepackage{times,cite}

\newcommand{\hide}[1]{}
\newcommand{\bibhide}[1]{}

\def\s{\scriptstyle}
\def\ss{\scriptscriptstyle}

\newcommand{\bqe}{\begin{eqnarray}}
\newcommand{\eqe}{\end{eqnarray}}
\newcommand{\nn}{\nonumber}

\def\o{\over}
\def\C{{\s C}}

\def\pmicSf{P'({\scriptstyle C}^{\ss S}_i(m),t)}
\def\pmicCf{P'(\C_i^{\ss C}(m),t)}

\def\pxf{{P}'(\xi,t)}

\def\pxc{P_c(\xi,t)}

\def\pxSf{{P}'(\xi^{\ss \c}(m),t)}
\def\pxCf{{P}'({\xi}^{\ss \S}(m),t)}

\def\pti{\tilde P_i(t)}

\def\ptip{\tilde P'_i(t)}
\def\ptjp{\tilde P'_j(t)}
\def\ptkp{\tilde P'_k(t)}
\def\pticone{\tilde P^c_i(t+1)}

\def\fav{{\bar f}(t)}

\def\crossij{{\cal C}_{\ss C_iC_j}^{(1)}(m)}
\def\crossjl{{\cal C}_{\ss C_jC_k}^{(2)}(m)}

\def\hamijS{d^H_{\ss\S}(i,j)}
\def\hamijC{d^H_{\ss\c}(i,j)}
\def\hamilS{d^H_{\ss\S}(i,k)}
\def\hamilC{d^H_{\ss\c}(i,k)}

\def\e{{\rm e}}

\def\MM{{ M}}
\def\S{{\cal S}}
\def\c{{\cal C}}
 
\usepackage{calc}
\newcounter{hours}\newcounter{minutes}
\newcommand{\now}{ %
\setcounter{hours}{\time/60} %
\setcounter{minutes}{\time-\value{hours}*60} %
\ifnum\theminutes <10 \thehours:0\theminutes\else \thehours:\theminutes\fi}

\def\authorheading{C. R. Stephens}
\def\shorttitle{The RG and Genetic Dynamics}

\begin{document}
\pagerange{1}{9}

\title{The Renormalization Group and the Dynamics of Genetic Systems}

\author{Christopher R. Stephens\email{stephens@nuclecu.unam.mx}$^*$}
 {$^*$Instituto de Ciencias Nucleares\\ UNAM, Circuito Exterior\\ 
Mexico D.F. 04510\\  stephens@nuclecu.unam.mx\\}

\bigskip\bigskip

  \abstract{ 
    In this brief article I show how the notion of coarse graining and
    the Renormalization Group enter naturally in the dynamics of 
    genetic systems, in particular in the presence of recombination. 
    I show how the latter induces a dynamics wherein coarse grained and
    fine grained degrees of freedom are naturally linked as a function
    of time leading to a hierarchical dynamics that has a 
    Feynman-diagrammatic representation. I show how this coarse grained
    formulation can be exploited to obtain new results.  
    }


\section {Introduction}
\label{sec:intro}

The Renormalization Group (RG), in its many diverse guises, has proved
to be an immensely powerful and useful tool in the treatment of
systems with many degrees of freedom; with applications covering
a huge gamut, from relativistic quantum field theory to the asymptotics 
of differential equations. In this short article I discuss another
arena where the RG appears in a very natural way - genetic dynamics.

By genetic dynamics I mean the dynamics of string-, or tree-like objects
whose evolution is governed by a set of genetic operators, the most
common of which are selection, mutation and recombination. Selection 
and mutation have been extensively studied by physicists (see for example
\cite{drossel}). Recombination however remains relatively untouched, 
although it has been extensively studied in biology (see for example 
\cite{burger}). The chief areas
of interest are population genetics, and associated fields, and evolutionary
computation. In both fields of interest one may be dealing with many, many
degrees of freedom and hence the normal RG motivation of reducing degrees
of freedom is valid. For instance, a typical protein has $O(10^4)$ aminoacids.

Here, I will concentrate more on showing that recombination naturally 
induces a coarse graining and the subsequent dynamics possesses a 
hierarchical structure wherein genetic configurations are related in the
past to more coarse grained genetic ``building blocks''. I show that 
such coarse grainings have a RG structure and lead to new results and insights
that can be, and have been, profitably used - principally in evolutionary
computation. Although the flavour of this article is quite different 
many of the results and more details can be found in the following
articles \cite{stewael,stewael1,stewael2,stephens:2001:GECCO,alden}.

\section{Genetic Dynamics}
\label{sec:gendyn}

Consider the dynamics of a population 
${\cal M}(t)\equiv \{\C_i(t)\}\subset \cal G$ of strings of equal 
length\footnote{This is not a restriction. The extension to non-fixed
length strings and trees has been considered by Poli and collaborators 
in the context of Genetic Programming (see \cite{poli} for a recent 
exposition).}, $N$, where $\cal G$ is the configuration space of string states 
and $\{\C_i(t)\}$ is the set of ``genotypes'' present in the population 
at time $t$. For simplicity we will assume binary bits, though nothing we 
shall present depends on this fact. We may naturally represent $\cal G$ 
as a $N$-dimensional hypercube, a natural metric being the Hamming distance, 
strings associated with adjacent vertices being Hamming distance one apart. 
The $N$ string loci form a complete orthonormal basis for the hypercube. 

The population evolves in discrete time under the action of an 
evolution operator ${\cal H}$, the action of which depends on the specific 
``genetic'' operators involved. Here, we will consider the three canonical 
operators - selection, mutation and recombination - whereupon we have
\bqe
{\cal M}(t+1)={\cal H}(\{f_i\},\{{\cal M}(t)\},\{p_k\},t){\cal M}(t)
\label{abstractdyn}
\eqe
In this case ${\cal H}$ depends on the reproductive fitness landscape, 
$\{f_i\}$, the population $\{{\cal M}(t)\}$ and the set of parameters, 
$\{p_k\}$, that govern the other genetic operators; e.g. mutation and 
recombination probabilities. For selection $P_i(t+1)\equiv P'_i(t)=F_{ij}P_j(t)$,
where $F_{ij}$ is the fitness matrix and $P_i(t)$ is the probability of 
finding the string $\C_i$ at time $t$. One of the most widely used selection 
schemes is proportional selection where the dynamics is given by 
$F_{ij}=(f_i/\fav)\delta_{ij}$, where $\fav$ is the 
average population fitness. It is usually considered as a unary operator. 

Mutation is also a unary operator and, typically, is such that every string bit 
flips to its complement
with probability $p_m$ every generation. Recombination, in distinction, is a
binary operator (although higher cardinality can be considered) and is 
such that a ``child'' string is formed by taking a certain number of bits from
one ``parent'' string and the complement from another ``parent'' string. For 
example, one can form $1111$ from parents $1010$ and $0101$ by taking the
first and third bits from $1010$ and the complementary second and fourth bits
from $0101$. One may specify the bits taken from the first parent using
a recombination ``mask'', $m$. In the above example the recombination mask
is $1010$ which signifies take bits one and three from the first parent 
(specified by the position of the ones) and two and four from the second.

The resultant dynamical equation describing the evolution of the probability
distribution for this system is 
\bqe
P_i(t+1)={\cal P}_{ij}P_c^j(t)
\label{maseqtwo}
\eqe 
where $P_c^i(t)$ is the probability to find strings of type $\C_i$ after 
selection and crossover. 

The mutation matrix, $\cal P$, has matrix elements
${\cal P}_{ij}=p_m^{\s d^{\ss H}(i,j)}(1-p_m)^{\s N-d^{\ss H}(i,j)}$,
where $d^H_{ij}$ is the Hamming distance between the two strings. For mutation
Hamming distance is clearly a very natural metric.
Note that (\ref{maseqtwo}) also applies for a finite population if we 
interpret the left hand side of (\ref{maseqtwo}) as the expected proportion of 
genotype $\C_i$ to be found at $t+1$ while any $P_i(t)$ on the right 
hand side are to be considered as the actual proportions found at $t$.

Explicitly $P^i_c(t)$ is given by
\bqe
{P_c}_i(t)=P'_i(t)+{\sum_{\ss m=1 }^{\ss 2^N}}\lambda_{ijk}(m)P'_j(t)P'_k(t)
\label{eqnpc}
\eqe
where $\lambda_{ijk}(m)$ is an interaction term between strings, that depends
on the particular crossover mask $m$, and ${\sum_{\ss m=1 }^{\ss 2^N}}$ is 
the sum over all possible recombination masks, $m\in \MM$, $\MM$ being the space 
of masks. Generically, $\lambda_{ijk}$ can be divided
into two terms, one associated with string destruction, $\lambda_{ijk}^d$,
and the other, $\lambda_{ijk}^c$, associated with string construction. 
Taking as target the string $111$, an example of the former is $111+000
\rightarrow 110+001$ while for the latter an example is the inverse of 
this process. To write these processes more explicitly we denote the set of 
bits inherited by an offspring from parent $\C_j$ as $\S$ and the bits 
inherited from parent $\C_k$, i.e. the set $\C_k-\S$, by $\c$. Naturally, 
$\S$ and $\c$ both depend on the particular crossover mask chosen. Then,
\bqe
\lambda_{ijk}^d=-p_c(m)\delta_{ik}(1-\delta_{ij})\crossij=-p_c(m)\delta_{ik}
(1-\delta_{ij})\theta(\hamijS)\theta(\hamijC)
\label{thetas}
\eqe
and
\bqe
\lambda_{ijk}^c=p_c(m)(1-\delta_{ij})(1-\delta_{ik})\crossjl=\nn\\
{p_c(m)\o2}(1-\delta_{ij}-\delta_{ik}-\delta_{ij}\delta_{ik})[\delta(\hamijS)\delta(\hamilC)
+\delta(\hamijC)\delta(\hamilS)]
\label{deltas}
\eqe
where $p_c(m)$ is the probability to implement the mask $m$
and the coefficients $\crossij$ and $\crossjl$,  
represent the probabilities that, given that $\C_i$ 
was one of the parents, it is destroyed by the crossover process, and
the probability that given that neither parent was $\C_i$ it 
is created by recombination. $\hamijS$ is the Hamming distance between the 
strings $\C_i$ and $\C_j$ measured only over the set $\S$, with 
the other arguments in (\ref{thetas}) and (\ref{deltas}) being similarly 
defined. Note that $\crossij$ 
and $\crossjl$ are properties of the crossover process itself and therefore 
population independent. It is clear that for recombination Hamming 
distance is not a natural metric. For example, consider two parent strings
$1111111111$ and $0000000000$. A one-point crossover implemented between the
last two bits leads to offspring $1111111110$ and $0000000001$ which are 
Hamming distance one from the respective parents. An equally probable
crossover between the 
fifth and six bits however, leads to $1111100000$ and $0000011111$, which
are Hamming distance five away from the parents. For a given $i$, 
$\lambda_{ijk}$ is a 
$2^N$-dimensional matrix, but is very sparse, there being only $O(2^N)$ non-zero
elements. Thus, the microscopic representation is very inefficient, there 
being very few ways of creating a given target by recombination of strings.
The vast majority of string recombination events are neutral in that they
lead to no non-trivial interaction. 

The equations (\ref{maseqtwo}) and (\ref{eqnpc}) yield an exact expression 
for the probability distribution governing the evolution for arbitrary 
selection, mutation and crossover. It takes into account exactly the 
effects of destruction and construction of strings.

\section{Coarse-Grained Evolution Equations}
\label{sec:CGEE}

The dynamics of the previous section is described by $2^N$ coupled, non-linear
difference equations representing the microscopic degrees of freedom, i.e.
the strings themselves. In the absence of recombination, the equations are
essentially linear and, as is well known, the resulting selection/mutation
problem can be recast in the guise of a two-dimensional, inhomogeneous
statistical mechanics problem, where powerful techniques such as the transfer
matrix approach can be invoked. However, save in very simple problems, such as 
a linear fitness landscape, even this simpler problem is formidable.
Recombination adds an extra layer of complexity. Naturally, in such 
problems one always wishes to find the correct effective degrees of freedom
so as to be able to affect an effective reduction in the dimensionality
of the problem. Such reductions can sometimes come about in a relatively
trivial fashion, for instance, if there is an underlying symmetry that is 
preserved by the action of the genetic operators. This occurs for instance
with selection and the genotype-phenotype map. As fitness acts at the
phenotypic level then a natural coarse graining from genotypes to phenotypes
occurs. This symmetry is not necessarily preserved by the other genetic 
operators. As a concrete example consider a fitness landscape where the 
fitness is given by the number of ones on the string (a simple paramagnet).
In this case the dynamics can be rewritten in terms of the $N$ phenotypes
rather than $2^N$ genotypes. The equation of motion for selection only is then
\bqe
P_n(t+1)={n\over \bar n(t)}P_n(t)
\eqe
where we denote phenotypes by $n$, the number of ones, and $\bar n(t)$ is the
average number of ones in the population at time $t$. The solution of these $n$
difference equations is
\bqe
P_n(t)={n^tP_n(0)\over \sum_{n=0}^Nn^tP_n(0)}
\eqe
Another example is that of the Eigen model, where the fitness landscape is
degenerate for all genotypes except one, the master sequence. At the level of 
selection only, given that there are only two phenotypes, there is a reduction
in the size of the configuration space from $2^N$ to $2$, i.e. a reduction
in the number of degrees of freedom from $N$ to $1$. However, if we include in 
the effect of mutation we see there is an induced breaking of the 
genotype-phenotype symmetry due to the fact that strings close to the 
master sequence in Hamming distance have a higher ``effective'' fitness 
\cite{stephensvargas}. 

As mentioned in the previous section, the string representation for recombination
is very inefficient due to the sparsity of the interaction matrix. This
is an indication that strings are not the natural effective degrees of 
freedom for recombination. So what are? To form the string $111$ with a 
recombination mask $100$ one can join strings $111$, $110$, $101$ and $100$
with either $111$ or $011$. In other words, for the first parent the second
and third bit values are unimportant and for the second the first bit value is
unimportant. Thus, it is natural to coarse grain over those strings that give
rise to the desired target for a given mask. Such coarse-grained variables
are known as ``schemata'', which we will denote by $\xi_i$, and are equivalent 
to, for instance, ``block spins'' in traditional statistical mechanics 
RG applications. The marginal 
probability, $\pti$, represents the probability of finding the schema $\xi_i$ 
at time $t$. A specific schema is determined by summing over those bit 
positions that are not part of the schema. One may denote such a bit position 
by a $*$. Thus, $11*$ represents the two strings $111$ and $110$. The number
of definite bits of the schema defines its order, $N_2$, while the distance
between the outermost defining bits defines its length. Thus, $*11**0**$ has
$N_2=3$ and $l=5$.

As there exist $3^N$ possible schemata a full schemata basis is 
overcomplete as well as being non-orthonormal and corresponds to the space
of all possible blocked spins. However, the space of schemata is not the 
natural one for recombination as we shall see. If one 
picks arbitrarily a vertex in $\cal G$, associated with a string $\C_i$, one may
perform a linear coordinate transformation 
$\Lambda:{\cal G}\rightarrow \tilde{\cal G}$ 
to a basis consisting of all schemata that contain $\C_i$. For instance, for
two bits ${\cal G} = \{11,10,01,00\}$, while $\tilde{\cal G}=\{11,1*,*1,**\}$.
The invertible matrix $\Lambda$ is such that 
$\Lambda_{ij}=1$ $\iff$ $\C_j\in \xi_i$.
We denote the associated coordinate basis the Building Block basis (BBB). 
The BBB is complete but
clearly not orthonormal. Note that the vertex $\C_i$ by construction is a
fixed point of this transformation. Apart from the vertex $\C_i$, note that
points in $\tilde{\cal G}$ correspond to higher dimensional objects in $\cal G$.
For instance, $1*$ and $*1$ are one-planes in $\cal G$ while $**$ is the whole
space. In the BBB one finds  
\bqe
\pticone=\ptip+{\sum_{\ss m=1 }^{\ss 2^N}}\tilde\lambda_{ijk}(m)\ptjp\ptkp
\eqe
where $\tilde\lambda_{ijk}(m)=\Lambda_{ii'}\lambda_{i'j'k'}
\Lambda^{-1}_{jj'}\Lambda^{-1}_{kk'}$. $\tilde\lambda_{ijk}(m)$ has the property that
for a given mask only interactions between BBs that construct 
the target schema are non-zero. i.e. $\tilde\lambda_{ijk}(m)=0$, unless $k$ 
corresponds to a schema which is the complement of $j$ with respect to $i$.  
For example, for two bits, if we choose as vertex 
$11$, then $11$ may interact only with $**$, while $1*$ may interact only with
$*1$. In $\cal G$ this has the interesting interpretation that for a target
schema $\xi$ of dimensionality $(N-d)$ only geometric objects ``dual'' in the 
$d$-dimensional subspace of $\cal G$ that corresponds to $\xi$ may interact. 
i.e. a $k$-dimensional object recombines only with a $(N-d-k)$-dimensional 
object.
Additionally, a $(N-d)$-dimensional object may only be formed by the interaction
of higher dimensional objects. In this sense interaction is via the geometric
intersection of higher dimensional objects. For example, the point $11$ can
be formed by the intersection of the two lines $1*$ and $*1$. Similarly, 
$1111$ can be formed via intersection of the three-plane $1***$ with the line
$*111$ or via the intersection of the two two-planes $11**$ and $**11$.  

Given that the object
dual to a vertex is always the trivial schema, $*...*$, where all bits are 
coarse grained, and $P(*...*,t)=1$, then it is instructive to combine the 
term linear in $P'_i(t)$ with it's pure selection counterpart. One obtains
for an arbitrary string $\C_i$ 
\begin{eqnarray}
P_c(\C_i,t)=(1-p_c)P'(\C_i,t)
+{\sum_{\ss m=1}^{\ss 2^N}}p_c(m)
\pmicSf\pmicCf
\label{stringfin}
\end{eqnarray}
where $p_c=\sum_mp_c(m)$ and we have returned to a less abstract notation.
$\pmicSf$ is the 
probability to select the BB $\C_i^{\ss\cal S}(m)$ and $\pmicCf$ the probability
to select the BB $\C_i^{\ss\cal C}(m)$. Both $\C_i^{\ss\cal S}(m)$
and $\C_i^{\ss\cal C}(m)$ are elements of the BBB. The above equation clearly
shows that recombination is most naturally considered in terms of the BBB. The
$(2^N-1)$ destruction terms associated with $\lambda_{ijk}^d$ have been 
reduced to only one term while the $(2^N-1)^2$ construction terms have also
been reduced to one term. Of course, we must remember that the coarse grained
averages of $\C_i^{\ss\cal C}(m)$ and $\C_i^{\ss\cal S}(m)$ contain $2^N$ 
terms, still, the reduction in complication is enormous. Thus, we see that
recombination as an operator naturally introduces the idea of a coarse graining.

Inserting (\ref{stringfin}) in (\ref{maseqtwo}) we can then try to solve 
for the dynamics. However, in order to do that we must know the time dependence
of the BB schemata $\C_i^{\ss\cal C}(m)$ and $\C_i^{\ss\cal S}(m)$. Although
the number of BB basis elements is $2^N$ we may generalize and consider the
evolution of an arbitrary schema, $\xi$. 
To do this we need to sum with 
$\sum_{\C_i\supset\xi}$ on both sides of the equation (\ref{maseqtwo}). 
This can simply be done to obtain \cite{stewael,stewael1,stewael2}
again the form (\ref{maseqtwo}), where this time the index $i$ runs only
over the $2^{N_2}$ elements of the schema partition and where again
${\cal P}_{ij}=p_m^{\s d^{\ss H}(i,j)}(1-p_m)^{\s N-d^{\ss H}(i,j)}$. In this
case however $d^H_{ij}$ is the Hamming distance between the two schemata.
For instance, for three bit strings the schemata partition associated with
the first and third bits is $\{1*1,1*0,0*1,0*0\}$. In this case $d^H_{12}=1$
and  $d^H_{14}=2$. $P_c(\xi,t)=\sum_{\C_i\supset\xi}P_c(\C_i,t)$ is the 
probability of finding the schema $\xi$ after selection and crossover. 
Note the form invariance of the equation after coarse graining. To complete 
the transformation to schema dynamics we
need the schema analog of (\ref{stringfin}). This also can be obtained by
acting with $\sum_{\C_i\supset\xi}$ on both sides of the equation. One obtains
\bqe
\pxc= (1-p_c{N_{\ss{\MM}_r}(\xi)\over N_{\ss\MM}})\pxf 
+ {\sum_{\ss m\in {\MM}_r}}p_c(m)\pxSf\pxCf\label{schemafin}
\eqe
where $\xi^{\ss \S}(m)$ represents the part of the schema $\xi$ inherited 
from the first parent and $\xi^{\ss \c}(m)$ that part inherited from the second. 
$N_{\ss{\MM}_r}(\xi)$ is the number of crossover masks that affect 
$\xi$, ${\MM}_r$ being the set of such masks. $N_{\ss\MM}$ is the total number
of masks with $p_c(m)\neq0$. Obviously, these quantities depend on the type 
of crossover implemented and on properties of the schema such as defining
length. 
 
Thus, we see that the evolution equation for schemata is form invariant
there being only a simple multiplicative renormalization of the crossover
probability $p_c$. This form invariance, first shown in \cite{stewael}, demonstrates
that BB schemata in general are a preferred set of coarse grained variables
and more particularly the BBB is a preferred basis in the presence of 
recombination. It has also been shown \cite{vose} that schemata, more generally, 
are the only coarse graining that leads to invariance in the presence of mutation 
and recombination.

Considering again the structure of (\ref{stringfin}) and (\ref{schemafin})
we see that variables associated with a certain degree of coarse graining are 
related to BB ``precursors'' at an earlier time, which in their turn ... etc. 
This hierarchical structure terminates at order one BBs as these are unaffected
by crossover. Thus, for example, the level one BB combinations of $111$, 
i.e. BBs that lead directly upon recombination to $111$ are: $11*:**1$, $1*1:*1*$
and $1**:*11$. The level two BBs are $1**$, $*1*$ and $**1$. Thus a typical 
construction process is that BBs $1**$ and $*1*$ recombine at $t=t_1$ to form 
the BB $11*$ which at some later time $t_2$ recombines with the BB $**1$ to form
the string $111$.

\section{Renormalization Group}
\label{sec:RG}

In the previous section we saw that coarse grained variables arise very 
naturally in genetic dynamics and gave as examples the genotype-phenotype map
and schemata. We can formalize these considerations by formally introducing a 
general coarse graining operator ${\cal R}(\eta,\eta')$ which coarse
grains from the variable $\eta$ to the variable $\eta'$. In this case
\bqe
{\cal R}(\eta,\eta')P(\eta,t)=P(\eta',t)\qquad\qquad\qquad
{\cal R}(\eta,\eta'')P(\eta,t)=P(\eta'',t)
\eqe
However, given that ${\cal R}(\eta',\eta'')P(\eta',t)=P(\eta'',t)$ we deduce
that 
\bqe
{\cal R}(\eta,\eta'')={\cal R}(\eta,\eta'){\cal R}(\eta',\eta'')
\eqe
i.e. the space of coarse grainings has a semi-group structure. Thus, we 
see that one can naturally introduce the RG into the 
study of genetic dynamics. The naturalness of a particular RG transformation
will be to a large extent determined by how the dynamics looks under this 
coarse graining. Considering (\ref{abstractdyn}) for the pdf of the dynamics
then given that ${\cal R}(\eta,\eta')P(\eta,t)=P(\eta',t)$ the dynamics under
a coarse graining is governed by ${\cal R}(\eta,\eta'){\cal H}_{\eta}P(\eta,t)$,
where ${\cal H}_{\eta}$ is the dynamical operator associated with the variables
$\eta$. If this can be written in the form ${\cal H}_{\eta'}P(\eta',t)$ with
suitable renormalizations then the dynamics is form covariant or invariant 
under this coarse graining. As we have seen, for selection only the dynamics
is invariant\footnote{In this case it is strictly invariant not just form 
invariant as there is no renormalization necessary of any parameter or variable.}
when passing from genotypic to phenotypic variables, while for schemata the 
whole dynamics is form invariant, although there is a non-trivial renormalization
of the fitness landscape as well as a simple renormalization of the
recombination probability. In the case of recombination note also that the 
coarse graining operator associated with the BBs satisfies
\bqe
{\cal R}(\eta,\eta')={\cal R}(\eta^{\ss\S},\eta'^{\ss\S})
{\cal R}(\eta^{\ss\c},\eta'^{\ss\c})
\eqe
where ${\cal R}(\eta^{\ss\S},\eta'^{\ss\S})$ represents the action of the
coarse graining on the BB $\S$ while ${\cal R}(\eta^{\ss\c},\eta'^{\ss\c})$ 
represents the action on the BB $\c$.

\section{Results from the coarse grained formalism}
\label{sec:RGresults}

One of the strengths of the present coarse grained formulation is that
much can be deduced simply by inspection of the hierarchical structure of the basic 
formulas. Introducing a $2^N$-dimensional population vector, ${\bf P}(t)$,
whose elements are $P(\C_i,t)$, $i=1,...,2^N$, equation (\ref{maseqtwo})
can then be written in the form
\bqe
{\bf P}(t+1)={\overline{\bf W}}_s(t){\bf P}(t)
+\sum_{m=1}^{2^N}p_c(m){\overline{\bf W}}{\bf j}(m,t)\label{matrixeqn}
\eqe
where the selection-crossover destruction-mutation matrix 
${\overline{\bf W}}_s(t)={\overline{\bf {\cal P}}}{\overline{\bf F}}(t)$.
The selection - crossover destruction matrix, ${\overline{\bf F}}(t)$, 
is diagonal, and takes into account selection and the destructive component 
of crossover. Explicitly, for proportional selection 
${\overline{\bf F}}_{ii}(t)=(f(\C_i)/\fav)(1-p_c)$. Finally, the ``source'' 
matrix is given by ${\bf j}(m,t)=\pmicSf\pmicCf$.
The interpretation of this equation is that ${\bf j}(m,t)$ is a source 
which creates strings (or schemata) by bringing BBs together,
while the first term on the right hand side tells us how the strings
themselves are propagated into the next generation, the destructive
effect of crossover renormalizing the fitness of the strings. 

As shown in \cite{stewael,stewael1,stewael2}, compared to a representation 
based on (\ref{eqnpc}),
even a formal solution of (\ref{stringfin}) in the absence of mutation and
for 1-point crossover and proportional selection yields much valuable 
qualitative information, such as a simple proof of Geiringer's theorem 
and an extension of it to the weak selection regime. Explicitly, we have
\bqe
{\bf P}(t)=\prod_{n=0}^{t-1}{\overline{\bf W}}_s(n){\bf P}(0)
+\sum_{m=1}^{2^N}p_c(m)\sum_{n=0}^{t-1}\prod_{i=n+1}^{t-1}{\overline{\bf W}}_s(i)
{\overline{\cal P}}\ {\bf j}(m,n)\label{solnnodiag}
\eqe
Due to the form invariance of the equations this solution actually 
holds true for arbitrary schemata. The only changes
are that the vectors are of dimension $2^{N_2}$, the matrices of dimension
$2^{N_2}\times2^{N_2}$, the sum over masks for the construction terms
is only over the set $\MM_r$ and that the BBs in $j(m,t)$ are
those of the schema rather than the entire string.  

The interpretation of (\ref{solnnodiag}) follows naturally from
that of (\ref{matrixeqn}). Considering first the case without mutation, 
the first term on the right hand side gives us the probability for propagating
a string or schema from $t=0$ to $t$ without being destroyed by crossover.
In other words $\prod_{n=0}^{t-1}{\overline{\bf W}}_s(n)$ is the Greens function
or propagator for ${\bf P}$. In the second term, ${\bf j}(m,n)$, 
each element is associated with the creation of a string or schema at time $n$
via the juxtaposition of two BBs associated with a mask $m$. 
The factor $\prod_{i=n}^{t-1}{\overline{\bf W}}_s(i)$ is the probability 
to propagate the resultant string or schema without crossover destruction 
from its creation at time $n$ to $t$. The sum over masks and $n$ is simply
the sum over all possible creation events in the dynamics. 

This formulation lends itself very naturally to a diagrammatic representation 
and the formulation of a set of ``Feynman rules'' which allow for the 
calculation of $P(\xi,t)$ in the BBB. Here for simplicity and transparency
we write them for $p_m=0$. The generalization to ``matrix'' propagators 
in the presence of mutation is straightforward. 
\vskip 0.1truein
\par\noindent 1) Draw all possible connected tree diagrams that contribute 
to $\xi$
\par\noindent
2) For each diagram to each internal line attach a propagator 

\bqe
{F}_{ij}(t,t')=(1-p_c{N_{\ss{\MM}_r}(\xi_j)\over N_{\ss\MM}})^{t-t'}
\prod_{n=t'}^t\delta_{ij}{f(\xi_j,n)\over \bar f(n)}\nn
\eqe
\par\noindent
3) To each vertex assign a weight

\bqe
\tilde\lambda_{ijk}=p_c(m){f(\xi^{\ss\S}_j(m),n)\over \bar f(n)}
{f(\xi^{\ss\c}_k(m),n)\over \bar f(n)}\delta(d_k+d_i-(N-d_j))\nn
\eqe
\par\noindent
4) Carry out the integration over time for all vertices 
\vskip 0.1truein

In the above $d_i$ represents the dimensionality of the schema $i$. 
As a simple example consider two bit strings and $p_c=1$. Consider 
the pdf for $11$. In this case there are only two diagrams; the diagram 
corresponding to propagation of $11$ itself from $t=0$ to $t$ and 
the formation of $11$ by recombination of $1*$ and $*1$ at time $t'$, 
$0\leq t' < t$. In this case $F_{11}(t,0)=0$,
$F_{1*}(t,0)=\prod_{n=0}^{t-1}(f(1*,n)/\bar f(n))$ and similarly for $*1$. 
For the interactions the only non-zero vertex is 
$\tilde\lambda_{\ss 11,1*,*1}=(f(1*,n)/\bar f(n))(f(*1,n)/\bar f(n))$. 
The diagrammatic series is naturally a perturbation series in the number
of BB recombination events. In the case of a flat fitness landscape
the entire diagrammatic series can be exactly resummed for an arbitrary string
in the continuous time limit to find \cite{stephens:2001:GECCO} 
\bqe
P(\C_i,t)=\sum_{n=0}^{N-1}\e^{-{np_ct\o N-1}}(1-\e^{-{p_ct\o N-1}})^{N-n-1}
{\cal P}(n+1)\label{exactsol}
\eqe
where ${\cal P}(n+1)$ is an initial condition and represents a partition over
the probabilities for finding $(N-n)$ building blocks at $t=0$. For a given 
$n$ there are ${}^{\ss N-1}C_n$ such terms. For instance, for $11$, 
${\cal P}(2)=P(11,0)$ and ${\cal P}(1)=P(1*,0)P(*1,0)$. 
Note that the simple dynamical form
arises because of the use of the BBB. One can use $\Lambda$ to rewrite the
above result in a string basis. The resulting expression is far more 
complicated in that the dynamical factors for a given string combination are
complicated combinations of those associated with the BBB. Note also that 
the fixed point is non-perturbative in $p_c$ indicating that the asymptotic
dynamics cannot be accessed perturbatively. This is why it was necessary to 
sum the entire diagrammatic series.

The tendency of recombination is to destroy correlations between different 
loci in the population. Selection, depending on the landscape, can induce
corrlations, hence there is a competition. In the case of weak selection and
strong crossover (\ref{exactsol}) shows that correlations asymptotically decay
and hence the effective degrees of freedom are $1$-schemata. Higher order 
schemata can be taken as perturbations around this decorrelated limit. In the
case of a recombination operator that mixes freely all bits within the entire
population (genepool recombination) then these perturbations are zero and
the $1$-schemata give an exact description of the dynamics. Under these 
circumstances one may solve the dynamics exactly, including mutation, for 
certain fitness landscapes, such as a linear fitness landscape \cite{alden}.

\section{Conclusions}
\label{sec:conclusions}

We have here briefly tried to lay out why RG concepts and techniques
can be useful when studying the dynamics of genetic systems. I showed 
that genetic dynamics can be profitable studied in the context
of coarse grained degrees of freedom. What is a natural coarse graining
was seen to depend on the genetic operators present. We saw that 
for recombination the equations of motion were form invariant under
a schemata coarse graining and that the BBB was a preferred one leading
to a recombination dynamics where effective degrees of freedom are related to
more coarse grained BBs. We showed that the hierarchical nature of
recombination led to a natural formulation in terms of Feynman diagrams with
an associated set of Feynman rules. We also briefly mentioned some concrete
results that emerge naturally from a coarse grained formulation. We strongly
believe that the RG has an important role to play in developing a more 
quantitative understanding of the dynamics of genetic systems and hope there
will be interesting devlopments to report at the next RG conference.  

\begin{ack}

This work was supported in by Conacyt grants 32729-E, 32399-E and
30422-E. I am grateful to Riccardo Poli, Alden Wright and Michael 
Vose for useful conversations and wish to thank the organizers of 
RG2002 for their very pleasant hospitality.

\end{ack}

\footnotesize


\begin{thebibliography}{99}



\bibitem{drossel}
\par\noindent Drossel, B. (2001) ``Biological Evolution and Statistical
Physics'', {\it Advances in Physics}, to be published; preprint 
cond-mat/0101409. 

\bibitem{burger}
\par\noindent B\"urger, R. (2000) ``The Mathematical Theory of Selection,
Recombination and Mutation'', Wiley Series in Mathematical and 
Computational Biology.

\bibitem{stewael} 
\par\noindent Stephens, C.R. and Waelbroeck, H. (1997), 
``Effective Degrees of Freedom in Genetic Algorithms and the Block
Hypothesis'', {\it Proceedings of the ICGA7\/}, ed. T. B\"ack, 34-41 
(Morgan Kaufmann, San Mateo).

\bibitem{stewael1} 
\par\noindent Stephens, C.R. and Waelbroeck, H. (1998) ``Analysis of the
Effective Degrees of Freedom in Genetic Algorithms'', {\it Physical Review}\ 
{\bf D57} 3251-3264. 

\bibitem{stewael2} 
\par\noindent Stephens, C.R. and Waelbroeck, H. (1999) ``Schemata Evolution
and Building Blocks'', {\it Evol. Comp.}\ {\bf 7(2)} 109-124.

\bibitem{stephens:2001:GECCO}
\par\noindent Stephens, C.R. (2001) ``Some Exact Results from a Coarse Grained
Formulation of Genetic Dynamics''. In L. Spector {\it et al} 
eds. {\it Proceedings of GECCO 2001\/}, 631-638 (Morgan Kaufmann, 
San Francisco).

\bibitem{alden}
\par\noindent Wright, Rowe, J., Poli, R. and Stephens, C.R. (2002) 
``A Fixed Point Analysis of a Genepool GA with Mutation'', 
accepted for publication (full paper) in {\it GECCO 2002}.

\bibitem{poli} 
\par\noindent Langdon, W. and Poli, R. (2001) ``Foundations of Genetic
Programming'', (Springer-Verlag).

\bibitem{stephensvargas}
\par\noindent Stephens, C.R., Mora, J. (2001)
``Effective Fitness as an Alternative Paradigm for Evolutionary Computation II:
Examples and Applications'', {\it Genetic Programming and 
Evolvable Hardware}, {\bf 2}, 7-32.

\bibitem{vose} 
\par\noindent Vose, M.D. (1999) {\it The Simple Genetic Algorithm: 
Foundations and Theory}, (MIT Press, Cambridge MA). 

\end{thebibliography}
\end{document}